\documentclass[aps,prd,twocolumn,showpacs,preprintnumbers,amsmath,amssymb,nofootinbib]{revtex4-1}
\usepackage{graphicx}
\usepackage{dcolumn}
\usepackage{bm}
\usepackage{slashed}
\usepackage{amsmath}
\usepackage{mathrsfs}

\newcommand{\ba}{\begin{array}}
\newcommand{\ea}{\end{array}}
\newcommand\gr[1]{\mathrm{#1}}    
\newcommand\muI{\mu_{\mathrm{I}}}
\newcommand\muIc{\mu_{\mathrm{I}}^{\mathrm{c}}}
\newcommand\muIcs{(\mu_{\mathrm{I}}^{\mathrm{c}})^2}
\newcommand\Nc{N_{\mathrm{c}}}

\newcommand{\ie}{{i.e.}}
\newcommand{\beq}{\begin{equation}}
\newcommand{\eeq}{\end{equation}}
\newcommand{\bea}{\begin{eqnarray}}
\newcommand{\eea}{\end{eqnarray}}
\DeclareMathOperator{\diag}{diag}
\begin{document}
\title{%
Fate of chiral critical point under the strong isospin asymmetry 
}%
\author{Hiroaki Abuki}%
\email[E-mail:~]{h.abuki@rs.tus.ac.jp}
\affiliation{%
Department of Physics, Tokyo University of Science, Tokyo 162-8601, Japan}%
\date{\today}
\begin{abstract}
We study the influence of the isospin asymmetry on the phase structure
 of strongly interacting quark matter near the critical point (CP)
 using a Ginzburg-Landau approach.
The effect is found to be drastic, not only bringing about the shift
 of the location of the CP, but resulting in a rich phase structure in
 the vicinity of the CP. 
In particular, new tricritical and triple points emerge as soon as the
 isospin density becomes finite.
Moreover, we find the CP being washed out from the phase diagram
 due to the stabilization of a homogeneous charged pion condensate
 when the isospin chemical potential exceeds a critical value.
We derive a model-independent universal relation between the critical
 isospin chemical potential and the chiral condensate at the CP.
We also study the effect of the $\mathrm{U}(1)_\mathrm{A}$ anomaly on
 the phase transition to the pion condensate in the vicinity of
 chiral crossover.
\end{abstract}
\maketitle

\section{Introduction}
The phase diagram of QCD at finite temperature and/or finite density is
 the subject of extensive theoretical and experimental studies. 
In particular, several approaches to QCD with two light flavors suggest
 the existence of a critical point (CP) at which the first-order chiral
 phase transition turns into a crossover \cite{Asakawa:1989bq}.
Despite many efforts based on the first principle calculations
 \cite{Fukushima:2010bq}, not only the precise location of the CP, but
 its existence itself remains controversial.

In the chiral limit with vanishing quark mass, the CP becomes a
 tricritical point (TCP).
The effect of a finite quark mass is, thus, rather simple; 
 just to smear the second-order chiral phase transition to a crossover
 and, accordingly turn the TCP into the CP. 
Our focus here is the other important ingredient in realistic systems,
 the effect of an isospin asymmetry. 
Such a flavor symmetry breaking is caused by a neutrality constraint
 that should be imposed in any bulk systems to prevent the diverging
 energy density. 

The isospin imbalance is known to bring a rich variety of color
 superconducting phases at high density \cite{Abuki:2004zk}. 
On the other hand, QCD at large isospin density was first studied in
 Ref.~\cite{Son:2000xc}, and it was shown that the QCD vacuum develops a
 pion condensate (PIC) as soon as $|\muI|>m_\pi$, where $m_\pi$ and
 $\muI$ are the vacuum pion mass and the isospin chemical potential,
 respectively.
The PIC can be viewed as a relativistic superfluid that exhibits
 a crossover from a Bose-Einstein condensate (BEC) of pions to a
 superfluidity of the Bardeen-Cooper-Schrieffer (BCS) type \cite{He:2005nk}.
So far, several model analyses have been made for the PIC at finite 
 temperature and/or quark density
 \cite{Klein:2003fy,Barducci:2004tt,Nishida:2003fb,%
 He:2005nk,He:2005tf,He:2006tn,Andersen:2007qv,%
 Abuki:2008wm,Abuki:2009hx}.
However, to our knowledge, there is at present, no systematic analysis
 based on the Ginzburg-Landau (GL) approach focusing on the isospin
 effects on the CP.
This is what we present here for the first time.

Our GL framework is advantageous to other approaches in the sense that
it can give model-independent predictions near the CP.
Since we are interested in the response of the CP and phases in its
 neighborhood against nonzero $\muI$, our strategy is to take $\muI$ as
 a perturbative field and expand the GL functional with respect to it.
We use a quark loop approximation to reduce the number of GL
 couplings.
This approximation should be valid, in particular, if it is located at
 a large fugacity region. 

In this paper, we restrict the analysis to homogeneous phases 
only, leaving more detailed analysis to future work \cite{Abuki}.
This is, to some extent, an extension of our previous work
 \cite{Iwata:2012bs} to the situation off the chiral limit 
 introducing a finite quark mass.
This paper is organized as follows. In Sec.~\ref{sec:fourth}, we derive
 a general GL potential up to the fourth order in fields and
 discuss the effects of $\muI$ and $\gr{U(1)_A}$ breaking at the
 vicinity of chiral crossover. 
In Sec.~\ref{sec:sixth}, we extend the GL potential up to the sixth
 order to discuss the isospin effect on the CP.
Based on this, we clarify how the CP and its neighborhood are affected
 by the inclusion of isospin asymmetry.
In Sec.~\ref{sec:conc}, we summarize.

\section{Ginzburg-Landau approach at fourth order}\label{sec:fourth}
Let us start with writing the most general GL potential for the two
 chiral four-vectors $\phi=(\sigma, \bm{\pi})$ 
 and its parity partner $\varphi=(\eta^\prime,\bm{a})$ with
 $\sigma\sim\langle\bar{q}q\rangle$, 
 $\bm{\pi}\sim\langle \bar{q}i\gamma^5\bm{\tau}q\rangle$,
 $\eta^{\prime}\sim\langle\bar{q}i\gamma_5 q\rangle$, 
 and $\bm{a}\sim\langle\bar{q}\bm{\tau}q\rangle$. 
At the fourth order in $\phi$ and $\varphi$, the chiral [%
 $\mathrm{SU}(2)_\mathrm{L}\times\mathrm{SU}(2)_{\mathrm{R}}\sim\mathrm{O}(4)$%
 ] symmetric part of the GL potential should take the form
\beq
\ba{rcl}
\Omega_0[\phi,\varphi]%
&=&\frac{1}{2}\alpha_2\phi^2+\frac{1}{2}%
\alpha_2^\prime\varphi^2\\[2ex]
&&+\frac{1}{4}\alpha_4\left((\phi^2+\varphi^2)^2%
+4(\phi^2\varphi^2-(\phi,\varphi)^2)\right),
\ea
\notag
\eeq
 where $(\phi,\varphi)\equiv \sigma\eta^\prime+\bm{a}\cdot\bm{\pi}$ is
 the inner product.
If $\alpha_2=\alpha_2^\prime$, the potential possesses the additional
 $\mathrm{U}(1)_\mathrm{A}$ symmetry, which is actually violated in QCD
 via the axial anomaly, so, typically, $\alpha_2\ne\alpha_2^\prime$.
The current quark mass and the isospin chemical potential add
 $\mathrm{O}(4)$ noninvariant terms to the potential. 
At the second order in fields,
\beq
\delta\Omega_{\mathrm{SB}}%
=-h\sigma+\beta_1(\sigma a_3+\eta^{\prime}\pi_3)+\beta_2\bm{\pi}_\perp^2,
\eeq
 where $\bm{\pi}_\perp=(\pi_1,\pi_2)$ is the charged pion doublet.
The first term is due to the current quark mass, and it explicitly
 breaks $\mathrm{O}(4)$ down to
 $\mathrm{SU}(2)_\mathrm{V}\sim\mathrm{O}(3)$.
The second and third terms are due to the finite isospin density, which
 violate the isospin $\mathrm{O}(3)$ into
 $\mathrm{U}_{\mathrm{I}_3}(1)\sim\mathrm{O}(2)$, 
 the rotation about the third axis of isospin space.
The GL coupling $h$ is proportional to the quark mass $m$ for light
 flavors. 
On the other hand, since the operator in the second (third) term is
 even (odd) under the isospin flip $u\leftrightarrow d$, we have
 $\beta_1\propto \muI$ and $\beta_2\propto\muI^2$ at the leading order
 in expansion in $\muI$.
When $\muI\ne0$, and $h\ne0$, we may anticipate the realization of
 following two phases:

\vspace*{0.5ex}
\noindent
{\bf (i) The chiral symmetry broken phase ($\chi$SB)}: 
the phase with $\sigma\ne 0$, which might be accompanied by a nonvanishing
 $a_3$.
The residual symmetry is the isospin, $\mathrm{O}(3)$.

\vspace*{0.5ex}
\noindent
{\bf (ii) The phase with a pion condensate (PIC)}: 
the phase with the charged pion condensate; $|\bm{\pi}_\perp|\ne0$. 
The $\gr{O(3)}$ symmetry is spontaneously broken down to
 $\mathrm{O}(2)$.

\vspace*{0.5ex}

We now have six GL couplings
 $\{\alpha_2,\alpha_2^\prime,\alpha_4,h,\beta_1,\beta_2\}$, and they are
 functions of thermodynamic variables $\{\mu,\muI,T\}$.
We can, in principle, investigate the phase structure in this 
 six-dimensional space in full generality.
However, even if we do that, it would become difficult to relate it with
the phase structure in the physical parameter space.
Instead of doing that, here we take the advantage of quark loop
 approximation for which we only take into account quark loop effects in
 the effective potential. 
This would give a reasonable approximation to the real potential in the
 high fugacity region.
The feedback of quark loops to the potential is
\beq
\Delta\Omega=-\frac{T\Nc}{V}
\sum_{n=2,4,\cdots}\frac{1}{n}\mathrm{Tr}\left(S_0 \Sigma\right)^n,
\notag
\eeq
 where $V$ denotes the spatial volume and $\gr{Tr}$ should be taken over
 the Dirac, flavor, and functional indices. 
$S_0 = \diag{(S_u, S_d)}$ is the bare quark propagator, and
 $\Sigma=\sigma{\bf{1}}+a_3\tau_3+i\gamma^5\pi\tau^1$ is the self-energy
 for which we set $\pi_2=0$ without any loss of generality. 
From this we can extract the following explicit expression for
 $\alpha_4$ in the leading order of expansion in $\muI$
 \cite{Nickel:2009ke}:
\beq
\alpha_{4}=\alpha_4^{(0)}(\mu,T)+{\mathcal O}(\muI^2),
\eeq
 where we have defined the quantity $\alpha_{2n}^{(0)}$ for $n\ge 1$ as
\beq
\alpha_{2n}^{(0)}(\mu,T)%
\equiv 8TN_c\sum_{n,{\textbf p}}\frac{1}{\left((i\omega_n+\mu)^2%
-{\textbf p}^2\right)^n},
\eeq
 with $\omega_n$ being the fermionic Matsubara frequency. 
Performing the series expansion in $\muI$ and discarding parts of
  integrand containing a total derivative in ${\textbf p}$
  \cite{Nickel:2009ke}, we can relate $\beta_2$ and $\beta_1$ with
  $\alpha_n^{(0)}$,
\beq
\ba{rcl}
 \beta_1%
  &=&\frac{1}{2}\mu_I\frac{\partial\alpha_2^{(0)}(\mu,T)}{\partial\mu}%
  +{\mathcal O}(\muI^3),\\[1ex]
 \beta_2%
  &=&-\frac{1}{4}\muI^2\alpha_4^{(0)}(\mu,T)+{\mathcal
 O}(\muI^4).\\[1ex]
\ea
\eeq
We introduce the GL parameter $\lambda$ by
 $(\beta_1/\muI)|_{\muI=0}=\lambda\alpha_4^{(0)}$; then we have
\beq
 \lambda=\frac{1}{2\alpha_4^{(0)}}%
 \frac{\partial\alpha_2^{(0)}(\mu,T)}{\partial\mu},
\eeq
which we use instead of $(\beta_1/\muI)|_{\muI=0}$ in the following.
In order to find explicit expressions for $\alpha_2$ and $\alpha_2^\prime$
 we need to specify the model. 
Let us consider here for a while the model with four-fermion interaction
 of the type
\beq
\ba{rcl}
{\mathcal L}_{\mathrm{int}}%
&=&\frac{G}{2}\left((\bar{q}q)^2+(\bar{q}i\gamma_5\bm{\tau}q)^2%
 +(\bar{q}i\gamma_5q)^2+(\bar{q}\bm{\tau}q)^2\right)\\[1ex]
&&+\frac{K}{2}\left((\bar{q}q)^2+(\bar{q}i\gamma_5\bm{\tau}q)^2%
 -(\bar{q}i\gamma_5q)^2-(\bar{q}\bm{\tau}q)^2\right).
\ea
\eeq
The interaction in the second line comes from so-called 
 Kobayashi-Maskawa--'t Hooft determinant, which violates the axial
 $\gr{U}(1)_\gr{A}$ symmetry explicitly. 
In this model, we have
 $\sigma=-(G+K)\langle\bar{q}q\rangle$,
 $a_3=-(G-K)\langle\bar{q}\tau_3q\rangle$, and
\beq
\ba{rcl}
\alpha_2&=&\frac{1}{G+K}+\alpha_2^{(0)}(\mu,T)+{\mathcal O}(\muI^6),\\[1ex]
\alpha_2^\prime&=&\frac{1}{G-K}+\alpha_2^{(0)}(\mu,T)+{\mathcal
 O}(\muI^6).
\ea
\label{eq:alpha2}
\eeq
We consider the following two cases here:

\vspace*{0.5ex}
\noindent
{\bf Case (I):}~%
Strong $\gr{U}(1)_{\mathrm{A}}$ breaking with $K=G$.
This corresponds to the standard Nambu--Jona-Lasinio (NJL) model with
\beq
{\mathcal L}_{\mathrm{int}}%
=G((\bar{q}q)^2+(\bar{q}i\gamma_5\bm{\tau}q)^2).
\label{eq:NJLtypica}
\eeq 
In this case, $\alpha_2^\prime$ diverges so that one of the chiral
 four-vectors $\varphi$ becomes irrelevant. 
In particular, $a_3=0$ \cite{Frank:2003ve,He:2005tf}.

\vspace*{0.5ex}
\noindent
{\bf Case (II):}~%
$\gr{U}(1)_{\mathrm{A}}$ symmetric case with $K=0$. 
In this case, $\alpha_2=\alpha_2^\prime$ so that we see that the GL
 potential also possesses the symmetry as it should. 

\vspace*{0.5ex}
\noindent

For case I, we find up to the fourth order in fields and $\muI$
\beq
\ba{rcl}
 \Omega%
 &=&\frac{\alpha_2}{2}\phi^2+\frac{\alpha_4^{(0)}}{4}\phi^4-h\sigma%
 -\frac{\alpha_4^{(0)}}{4}\muI^2\bm{\pi}_\perp^2.
\ea
\eeq
Here, $\alpha_2$ should be regarded as one evaluated at $\muI=0$, \ie, it
 should be understood as the lowest order in expansion in $\muI$,
 Eq.~(\ref{eq:alpha2}).
If we introduce the notation
\beq
\sigma_u=\frac{\sigma+a_3}{2},\quad\sigma_d=\frac{\sigma-a_3}{2},
\eeq
then we are forced to have $\sigma_u=\sigma_d=\sigma/2$.

For case II, we have at the same order
\beq
\ba{rcl}
\Omega%
&=&\frac{\alpha_2}{2}(\phi^2+\varphi^2)\\[1ex]
&&+\frac{\alpha_4^{(0)}}{4}\left(%
 (\phi^2+\varphi^2)^2+4(\phi^2\varphi^2-(\phi,\varphi)^2)\right)\\[1ex]
&&-h\sigma+\frac{\alpha_4^{(0)}}{4}\muI^2\bm{\pi}_\perp^2%
 +\alpha_4^{(0)}\lambda\muI(\sigma a_3+\eta^\prime\pi_3).
\ea
\eeq
For stability, we can only investigate the region $\alpha_4>0$ at this
 order, and we have typically $\lambda\ge0$ for $\mu\ge0$.
$\sigma_u$ and $\sigma_d$ in this case are proportional to $\bar{u}u$
 and $\bar{d}d$ condensates.
We notice that the $\lambda$ term in the potential contains
$\lambda\alpha_4\muI(\sigma_u^2-\sigma_d^2)$ so that positive $\muI$
 favors the situation $\sigma_d>\sigma_u$.
In the following analysis, we suppress the subscript $(0)$ in couplings,
so we have $\muI$ and four independent parameters
 $\{\alpha_2,\alpha_4,h,\lambda\}$, which do not depend on $\muI$.

\begin{figure}[t]
\begin{center}
\includegraphics[width=70mm]{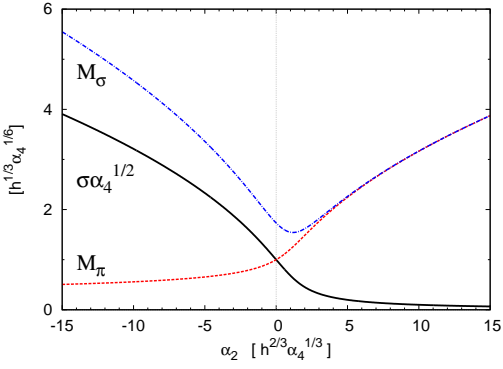}
\caption{The chiral condensate $\sigma$ (solid), the sigma mass
 $M_\sigma$ (dashed, blue), and the pion mass $M_\pi$ (dotted, red) as a
 function of $\alpha_2$, at $\muI=0$.
}
\label{fig:crossover}
\end{center}
\end{figure}

\begin{figure}[t]
\begin{center}
\includegraphics[width=70mm]{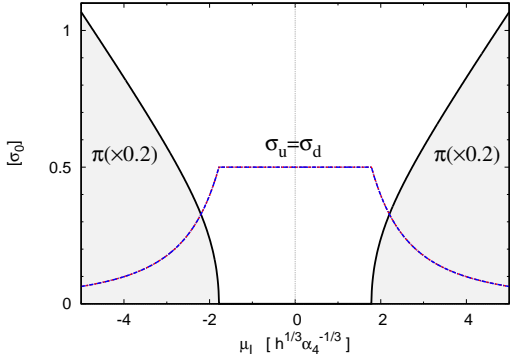}
\caption{The behavior of condensates as a function of $\muI$ at the
 crossover point, Eq.~(\ref{eq:puseudo}). 
$\sigma_u$ and $\sigma_d(=\sigma_u)$ are depicted by dashed lines (red
 and blue, respectively), while $\pi$ is shown by a solid line.
}
\label{fig:picon1}
\end{center}
\end{figure}

Let us first investigate case I. 
Introducing dimensionless parameters $\tilde{\sigma}$,
 $\tilde{\bm{\pi}}$, $\tilde{\muI}$, and $\tilde{\alpha_2}$ via
$$
\ba{l}
\sigma=\tilde{\sigma}\,(h/\alpha_4)^{1/3},\quad%
\bm{\pi}=\tilde{\bm{\pi}}\,(h/\alpha_4)^{1/3},\\[1ex]
\alpha_2\equiv \tilde{\alpha}_2\,h^{2/3}\alpha_4^{1/3},\quad%
\muI\equiv\tilde{\muI}\,(h/\alpha_4)^{1/3},
\ea
$$
 the potential is cast into
$
\Omega%
 =\alpha_4^{-1/3}h^{4/3}\omega
$,
where $\omega$ does not have any explicit dependence on $h$:
$$
\omega(\tilde{\sigma},\tilde{\bm{\pi}};\tilde{\alpha}_2,\tilde{\muI})%
 \equiv\frac{\tilde{\alpha}_2}{2}\tilde{\phi}^2+\frac{1}{4}\tilde{\phi}^4%
 -\tilde{\sigma}-\frac{1}{4}\tilde{\muI}^2\tilde{\bm{\pi}}_\perp^2,
$$
with $\tilde{\phi}\equiv(\tilde{\sigma},\tilde{\bm{\pi}})$.
When $h=0$ and $\muI=0$, the system has a second-order phase
 transition at $\alpha_2=0$. 
At finite $h\ne0$, the transition gets smoothed to a crossover.
In Fig.~\ref{fig:crossover}, we show the behavior of chiral condensate
 $\sigma$, sigma and pion masses ($M_\sigma$, $M_\pi$) as a function of
 $\alpha_2$. 
We find a crossover from the $\chi\gr{SB}$ to an approximately restored
 phase with $\sigma\sim0$ when $\alpha_2$ is increased.
We can define the pseudocritical point $\alpha_2=\alpha_2^{\gr{pc}}$,
for example, by the point where $M_\sigma$ takes the minimum. 
In this case, it can be numerically read as
\beq
 \alpha_2^{\gr{pc}}=1.191\,h^{2/3}\alpha_4^{1/3}.
\label{eq:puseudo}
\eeq
At this point, pion and sigma masses are
\beq
 M_\sigma^{\gr{pc}}=1.54\,h^{1/3}\alpha_4^{1/6},\quad
 M_\pi^{\gr{pc}}=1.26\,h^{1/3}\alpha_4^{1/6}.
\eeq
The ratio $M_\sigma^{\gr{pc}}/M_\pi^{\gr{pc}}=1.22$ at the
 pseudocritical point is universal to this order of GL expansion.
The chiral condensate $\sigma$ is also read as
\beq
 \sigma=0.630\,h^{1/3}\alpha_4^{-1/3}(\equiv\sigma_{\gr{pc}}).
\eeq
For what follows, we concentrate on the effect of $\muI$ at the
 pseudocritical point. 
In Fig.~\ref{fig:picon1}, we show $\sigma_u$, $\sigma_d$ and
 $|\bm{\pi}_\perp|\equiv\pi$ as a function of $\muI$. 
We see a second-order phase transition to the PIC phase at
\beq
|\muI|=1.782\,h^{1/3}\alpha_4^{-1/3}(\equiv\muI^\mathrm{c}).
\eeq
This can be written in terms of $\sigma_{\gr{pc}}$ or $M_\pi^{\gr{pc}}$ as
\beq
\muI^\mathrm{c}=2.83\,\sigma_{\gr{pc}}=(1.41/\sqrt{\alpha_4})\,M_\pi^{\gr{pc}}.
\eeq
We notice that what is in the universal relation in the GL framework at
 this order is the ratio of $\muI$ to the flavor singlet quark
 condensate $\sigma$ rather than that to the pion mass $M_\pi$.
Once $|\muI|\ge\muI^\mathrm{c}$, the charged pion condensate $\pi$
 develops.

A couple of questions are in order here:
i)~First, one might think that $\muI^{\mathrm{c}}$ determined here by
 looking at the static correlation function might be different from the
 true one, which should be determined by the pole of the charged pion
 propagator due to the kinetic seesaw mechanism
 \cite{Schafer:2001bq,Miransky:2001tw}.
ii)~Second, one might wonder what is the difference of $\pi^+$ and
 $\pi^-$ condensations.
Let us first discuss the point i. 
Actually, this is not the case, and the critical point determined by
 the static effective potential exactly coincides with the one by the
 charged pion propagators. 
If we worked out the time-derivative expansion in charged pion
 fields within the Gaussian approximation, we would have obtained
 the effective action density, which looks like (with a suitable
 normalization of $\pi$ fields)
\beq
\ba{rcl}
  {\mathcal L}_{\mathrm{eff}}&=&\mathrm{tr}%
  \big[(i\partial_t\bm{\pi}_\perp%
  +[\hat{\mu}_\mathrm{I},\bm{\pi}_\perp])^\dagger%
  (i\partial_t\bm{\pi}_\perp%
  +[\hat{\mu}_\mathrm{I},\bm{\pi}_\perp])\big]\\[1ex]
 &&-\frac{1}{2}M^2(\pi_1^2+\pi_2^2),
\ea
\eeq
where $M$ is the mass parameter being a function of
 microscopic/thermodynamic variables,
 $\bm{\pi}_\perp\equiv(\pi_1\tau_1+\pi_2\tau_2)/2$ and
 $\hat{\mu}_\mathrm{I}\equiv\muI\tau_3/2$, with
 $\{\tau_1,\tau_2,\tau_3\}$ being the Pauli matrices.
The charged pion propagator can be read as
\beq
\ba{rcl}
 D_{\pi_i\pi_j}^{-1}(\omega)
 &=&\begin{pmatrix}
 -\omega^2+M^2-\muI^2 & -2i\muI\omega\\
 +2i\muI\omega &-\omega^2+M^2-\muI^2\\
 \end{pmatrix}.\\[1ex]
\ea
\eeq
The static part of the propagator is related with the second derivative
 of effective potential:
\beq
 M^2-\muI^2\equiv%
 \frac{\partial^2\Omega}{\partial\pi_1^{\;2}}%
 \bigg|_{\bm{\pi}_\perp=0}%
 =\frac{\partial^2\Omega}{\partial\pi_2^{\;2}}%
 \bigg|_{\bm{\pi}_\perp=0}.
\eeq
The determinant of the polarization matrix includes all the pole masses:
\beq
 \det D_{\pi_i\pi_j}^{-1}=(\omega^2-M_+^2)%
 (\omega^2-M_-^2)
\eeq
$M_+\equiv M-\muI$ corresponds to the $\pi^+$ pole, while
$M_-\equiv M+\muI$ represents the $\pi^-$ pole.
The critical condition is given by the vanishing of either $\pi^+$
or $\pi^-$ mass, and, in both cases, $M^2-\muI^2=M_+M_-%
=\sqrt{\det D^{-1}_{\pi_i\pi_j}|_{\omega\to0}}=0$.
This is equivalent with the condition
\beq
  \frac{\partial^2\Omega}%
  {\partial\pi_1^{\;2}}\bigg|_{\bm{\pi}_\perp=0}%
  =\frac{\partial^2\Omega}%
  {\partial\pi_2^{\;2}}\bigg|_{\bm{\pi}_\perp=0}=0.
\eeq
This means that even if the pion masses split due to the kinetic 
  seesaw mechanism, the critical chemical point can be always
  obtained by looking at the behavior of static susceptibility.
Of course, we can only extract from $\Omega$ the multiple of two poles
  $M_+M_-$, not separately $M_+$ and $M_-$.
In other words, it is the geometric average of the pole masses that can
 be read from the curvature mass of $\Omega$.
Let us now come to point ii. 
Since the time dependence of the charged condensate is determined by
 the chemical potential as $\pi^+=\pi_1-i\pi_2\sim e^{-i\muI t}\pi$, 
 \ie, $(\pi_1,\pi_2)\sim(\pi\cos\muI t,\pi\sin\muI t)$,
 the rotation is counterclockwise when $\muI>0$ while it is clockwise
 for $\muI<0$.
We shall refer to the former as $\pi^+$ condensation, and the latter as
 $\pi^-$ condensation.

\begin{figure}[t]
\begin{center}
\includegraphics[width=70mm]{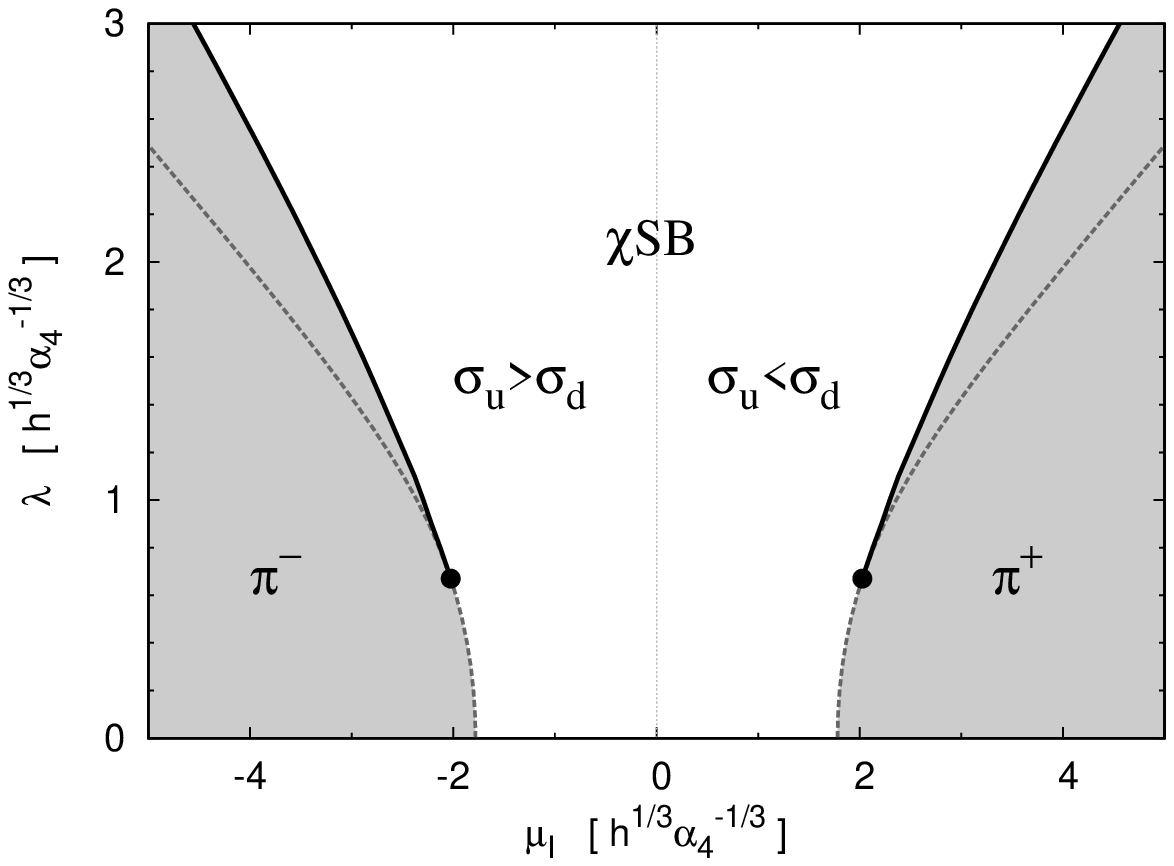}
\includegraphics[width=70mm]{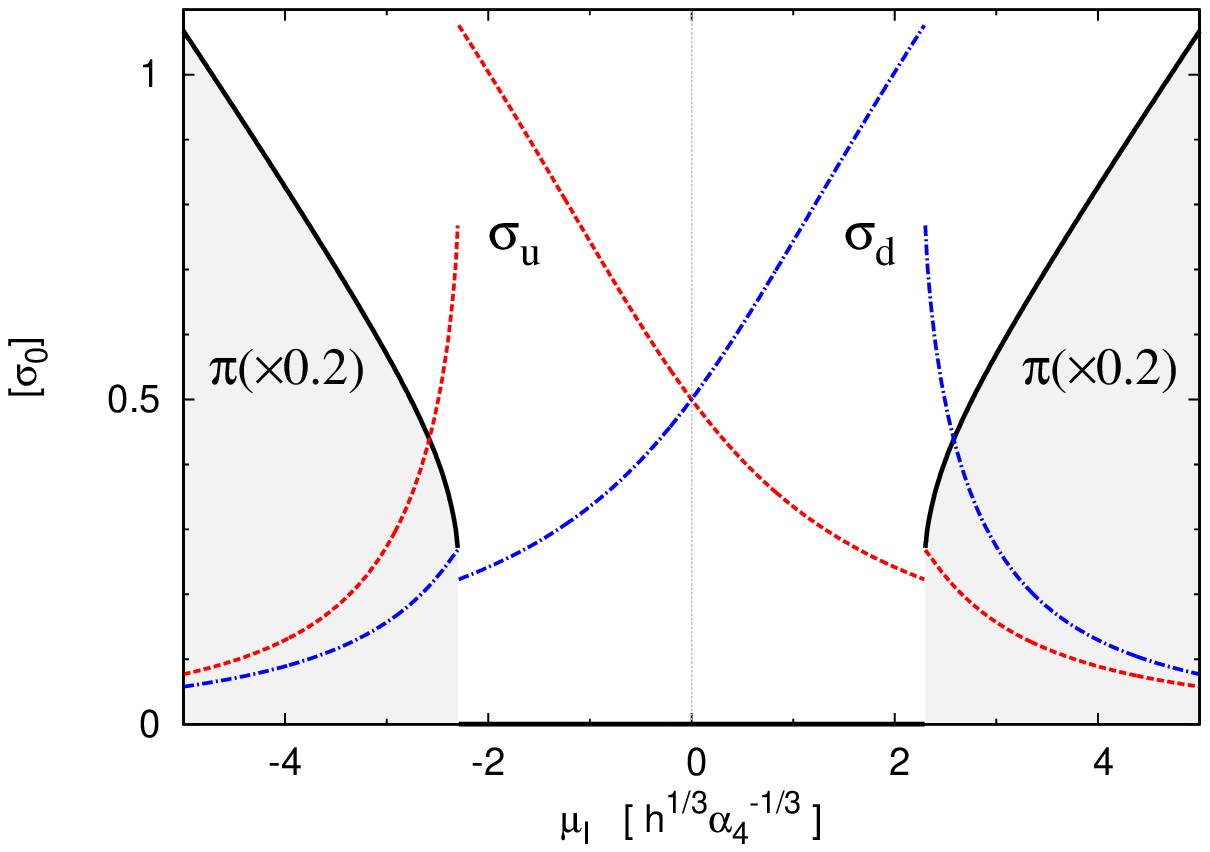}
\caption{(Upper panel):~The GL phase diagram in the
 $(\muI,\lambda)$ plane. 
 The solid line represents the first-order phase transition, while the
 dashed line represents second-order one separating the $\chi\gr{SB}$
 and PIC phases.
 The dashed line inside the shaded region expresses the spinodal line
 where the state without pion condensate becomes unstable.
(Lower panel):~The behavior of $\sigma_u$ (dashed, red), $\sigma_d$
 (dotted-dashed, blue), and $\pi$ (solid) as a function of $\muI$ 
 at $\lambda=1\,h^{1/3}\alpha_4^{-1/2}$.
}
\label{fig:phase1}
\end{center}
\end{figure}

Now, let us move on to case II. 
In this case, we have an additional GL parameter $\lambda\ge 0$ for
 $\mu\ge0$.
In the upper panel of Fig.~\ref{fig:phase1}, we display the phase
 diagram in the two-dimensional GL parameter
 space:~the $(\muI,\lambda)$ plane.
We have basically two phases, the $\chi\gr{SB}$ and the PIC with
 a $\pi^+$ or $\pi^-$ condensate.
However, in this case, we have $\sigma_u=\sigma_d$ only on two lines
 specified by $\muI=0$ or $\lambda=0$ and $\sigma_u\ne\sigma_d$ in the
 major part of the plane.
This is, of course, because of the $\lambda$ term in the GL potential,
 $\alpha_4\lambda\muI(\sigma_u^2-\sigma_d^2)$.
On the line $\lambda=0$, we have completely the same situation as
 displayed in Fig.~\ref{fig:picon1} in which 
 two second-order phase transitions are found at $\muI=\pm\muIc$.
On the other hand, when $\lambda$ becomes large, the transitions
 eventually change to the first-order ones. 
We show the situation at $\lambda=1\,[h^{1/3}\alpha_4^{-1/3}]$ in the
 lower panel of Fig.~\ref{fig:phase1}. 
We see clearly finite abrupt gaps in the order parameters associated
 with the first-order phase transition.
This is attributed to the competition between the $\lambda$ term and 
 the $\muI^2$ term in the potential;
 the former favors a larger $|\sigma_u-\sigma_d|$, while the
 latter likes the situation $\pi\ne0$ in which a smaller
 $|\sigma_u-\sigma_d|$ is favorable.
Accordingly, there are two tricritical points
 $(\muI,\lambda)=(\pm\muI^{\gr{TCP}},\lambda^{\gr{TCP}})$ at which the
 second-order phase transitions turn into first order ones.
Numerically, we find
\beq
\ba{rcl}
\muI^{\gr{TCP}}&=&2.02\,h^{1/3}\alpha_4^{-1/3}=3.21\,\sigma_{\gr{pc}}\\[1ex]
\lambda^{\gr{TCP}}&=&0.67\,h^{1/3}\alpha_4^{-1/3}=1.06\,\sigma_{\gr{pc}}.
\ea
\eeq
The ratio $\lambda^{\gr{TCP}}/\muI^{\gr{TCP}}=0.33$ does not depend on
 any of GL parameters and, thus, is universal to this order.
We note that the effect of strong flavor mixing due to the $\gr{U(1)_A}$
 anomaly makes the $\lambda$ term irrelevant via locking the
 condensates $\sigma_u$ and $\sigma_d$ with the same value
 \cite{He:2005tf,Frank:2003ve},  and, thus, renders the transition
 a second-order one.

\section{The Ginzburg-Landau approach at sixth order}\label{sec:sixth}
We now extend the GL analysis up to the sixth order so as to
 explore the influence of $\muI$ near the CP of QCD.
Doing this in full generality introduces many new GL parameters, which
 makes the analysis quite complicated.
Instead of doing this, we examine here only case I in the
 previous section, in which one of chiral four-vectors, $\varphi$,
 decouples.
The general GL for homogeneous condensates can be again written in terms
 of $\phi$ \cite{Iwata:2012bs}
\beq
\ba{rcl}
\Omega%
&=&-h\sigma+\frac{\alpha_2}{2}\phi^2-\beta_2\bm{\pi}_\perp^2\\[1ex]
&&+\frac{\alpha_4}{4}\phi^4+\frac{\beta_4}{4}\bm{\pi}_\perp^4%
 +\frac{\beta_{4b}}{4}(\phi^2-\bm{\pi}_\perp^2)\bm{\pi}_\perp^2%
 +\frac{\alpha_6}{6}\phi^6.
\ea
\eeq
GL coefficients $\alpha_n$ $(n=2,4,6)$ are expanded in the series of
 $\muI$ within the quark loop approximation (up to total derivatives) as
\beq
\begin{pmatrix}
 \alpha_2\\
 \alpha_4\\
 \alpha_6
\end{pmatrix}%
=\begin{pmatrix}
 1 & a\muI^2 & b\muI^4\\
 0 & 1 & c\muI^2 \\
 0 & 0 & 1 
\end{pmatrix}
\begin{pmatrix}
 \alpha_2(\muI=0)\\
 \alpha_4(\muI=0)\\
 \alpha_6(\muI=0)
\end{pmatrix}.
\eeq
Via explicit computations, we find $a=b=0$, and $c=1$.
Similarly, for $\beta_2$, $\beta_4$'s we find
\beq
\begin{pmatrix}
 \beta_2\\
 \{\beta_4,\beta_{4b}\}
\end{pmatrix}%
=\muI^2\begin{pmatrix}
 d & e\muI^2 \\
 0 & \{f,f_b\} 
\end{pmatrix}
\begin{pmatrix}
 \alpha_4(\muI=0)\\
 \alpha_6(\muI=0)
\end{pmatrix},
\eeq
 with $d=-{1}/{4}$, $e=0$, and $f=f_b=-2$.
Putting them all together, and assuming the condensate
 to have a charged pion component, \ie, $\phi=(\sigma,\pi,0,0)$,
 we arrive at
\beq
\ba{rcl}
\Omega%
&=&-h\sigma+\frac{\alpha_2}{2}\sigma^2%
 +\frac{\alpha_2-\muI^2\alpha_4/2}{2}\pi^2\\[1ex]
&&+\frac{\alpha_4+\muI^2\alpha_6}{4}\sigma^4%
 +\frac{\alpha_4+\muI^2\alpha_6/2}{2}\sigma^2\pi^2%
 +\frac{\alpha_4}{4}\pi^4\\[1ex]
&&+\frac{\alpha_6}{6}(\sigma^2+\pi^2)^3.
\ea
\label{eq:GLpot}
\eeq
Here $\alpha_2$, $\alpha_4$ and $\alpha_6$ should be understood as
 those evaluated at $\muI=0$ and, thus, are functions of $\mu$ and $T$
 only. 
$\alpha_6^{-1/2}$ has a dimension of energy so we use this as the unit
 of energy. 
In the following analysis, we, thus, set $\alpha_6=1$, but the proper
 dimension of any quantity can be recovered any time by use of
 $\alpha_6^{-1/2}$. 
Now, assuming $h>0$, we get rid of $h$ via scaling
\beq
\ba{rcl}
\sigma&=&\tilde{\sigma}\,h^{1/5},\quad\pi=\tilde{\pi}\,h^{1/5},%
\quad\muI=\tilde{\mu}_{\gr{I}}\,h^{1/5},\\[1ex]
\alpha_2&=&\tilde{\alpha}_2\,h^{4/5},\quad\alpha_4=\tilde{\alpha}_4\,h^{2/5}.
\ea
\eeq
Then, $h$ is scaled out of the potential as $\Omega=h^{6/5}\omega$.
Therefore, we now need to explore the GL phase diagram in the 
 three-dimensional GL parameter space $(\alpha_2,\alpha_4,\muI)$.

\begin{figure}[t]
\begin{center}
\includegraphics[width=80mm]{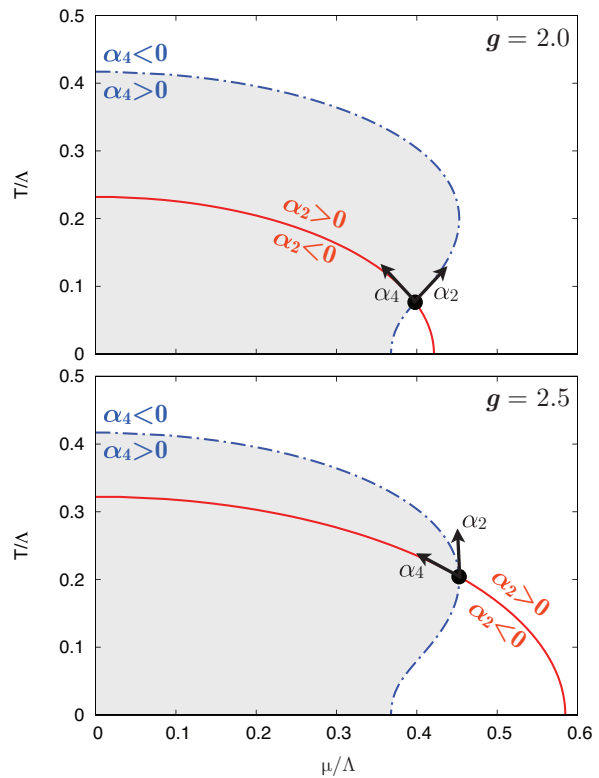}
\caption{The illustrative figure that shows how $(\alpha_2,\alpha_4)$
 spans the local coordinate in the vicinity of the TCP within the NJL
 model defined by Eq.~(\ref{eq:NJLtypica}) and the three-momentum cutoff
 $\Lambda$.
This is depicted for two values of coupling $g\equiv
 G\Lambda^2=2.0$ (upper panel) and $g=2.0$ (lower panel).
The solid line (red online) shows the curve on which $\alpha_2=0$, while
the dotted-dashed line (blue online) does that for $\alpha_4=0$.
The point of intersection gives the location of the TCP.
The region for $\alpha_4>0$ is shaded. 
The solid line in the shaded area represents the second-order chiral
 transition, while that in the unshaded area only gives the spinodal
 line on which the Wigner phase with $\sigma=0$ ceases to be a local
 minimum.
}
\label{fig:map}
\end{center}
\end{figure}

\begin{figure}[t]
\begin{center}
\includegraphics[width=0.35\textwidth]{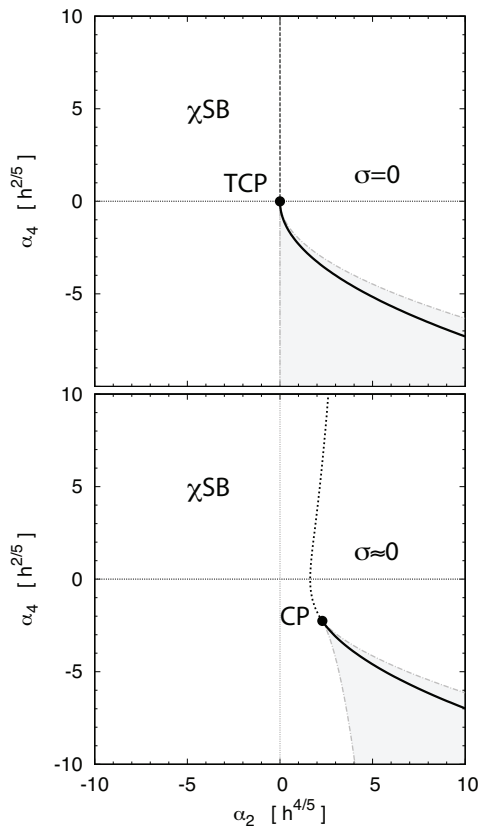}
\caption{The GL phase diagrams in the absence of $\muI$.
The phase diagram in the chiral limit $h=0$ (upper panel) and the one
 off the chiral limit (lower panel).
The shaded area represents the spinodal region. 
See the text for detail.
}
\label{fig:muI0}
\end{center}
\end{figure}

\vspace*{0.0ex}
\emph{How does $(\alpha_2,\alpha_4)$ map onto the $(\mu,T)$-plane?}~%
Before going into the discussion of GL phase diagrams, let us briefly
sketch how the GL parameters $(\alpha_2,\alpha_4)$ map onto the
$(\mu,T)$ plane taking the chiral limit $(h=0)$ for simplicity.
Figure~\ref{fig:map} illustrates how $(\alpha_2,\alpha_4)$ spans the
local coordinate in the NJL model specified by Eq.~(\ref{eq:NJLtypica})
and three-momentum cutoff $\Lambda$.
The upper panel shows the situation for $g=G\Lambda^2=2.5$, while the
lower panel shows that for $g=2.0$.
In the figure, the solid line expresses the curve determined by
$\alpha_2=0$, which separates the $(\mu,T)$-plane into two regions,
one for $\alpha_2>0$ and the other for $\alpha_2<0$.
Similarly, the dotted-dashed line shows the curve for $\alpha_4=0$.
The point of intersection determines the location of the TCP.
The region $\alpha_4>0$ is shaded just for a guide. 
The solid line inside this region determines the second-order chiral
phase transition, while that outside it only specifies the spinodal line
on which the Wigner phase ceases to be even a local minimum.
The axises of the local coordinate system $(\alpha_2,\alpha_4)$ are
depicted by arrows starting from the TCP; the $\alpha_2$ ($\alpha_4$)
coordinate points to the direction for $\alpha_2>0$ $(\alpha_4>0)$ side
with being tangent to the line of $\alpha_4=0$ ($\alpha_2=0$).
The location of the TCP as well as how $(\alpha_2,\alpha_4)$ maps onto
the $(\mu,T)$ plane depends on the detail of the model or regularization
scheme, so, for what follows, we only discuss the phase diagram in the
$(\alpha_2,\alpha_4)$ plane.

\vspace*{0.0ex}
\emph{GL phase diagram for quark matter without an isospin imbalance:~}
Let us start with the case with $\muI=h=0$. 
In the upper panel of Fig.~\ref{fig:muI0}, the phase diagram for this
 case is displayed.
In this case, $h$ in the figure labels can be regarded as an arbitrary
 unit; changing it does not modify the phase diagram.
There are two phases: the $\chi\gr{SB}$ phase with $\sigma\ne0$ and the
 symmetric (Wigner) phase. 
For $\alpha_4>0$, these phases are separated by a second-order phase
 transition located at $\alpha_2=0$, which is depicted by the dashed
 line. 
For $\alpha_4<0$, the transition is replaced by the first-order one at
 $\alpha_2=\frac{3}{16}\alpha_4^2$ shown by a solid line
 \cite{Nickel:2009ke}.
Accordingly, the TCP is located at the origin.
The shaded area shows a spinodal region in which one of the states exists
 as a local minimum of the potential; at $\alpha_2=0$ for $\alpha_4<0$,
 the Wigner phase becomes unstable, while at the line
 $\alpha_2=\frac{1}{4}\alpha_4^2$ for $\alpha_4<0$, a local minimum
 corresponding to the $\chi\gr{SB}$ state vanishes.
In the lower panel of Fig.~\ref{fig:muI0}, we show the phase diagram 
 for $h\ne0$. 
In this case, we have only the situation $\sigma\ne0$. 
Nevertheless, the first-order phase transition survives and separates 
 the $\chi\gr{SB}$ phase and a nearly symmetric phase with
 $\sigma\sim0$.
The first-order phase transition ends at the CP. 
The exact location of the CP, 
 $(\alpha_2^{\gr{CP}},\alpha_4^{\gr{CP}})$, 
 is derived analytically \cite{Friman:2012gg}
\beq
\textstyle
(\alpha_2^{\gr{CP}},\alpha_4^{\gr{CP}})%
 =\left(\frac{5}{4}\frac{3^{4/5}}{2^{2/5}}h^{4/5},%
 -\frac{5}{2^{1/5}3^{3/5}}h^{2/5}\right).
\label{eq:CP}
\eeq
This is numerically evaluated as $\sim(2.28h^{4/5},-2.25h^{2/5})$.
The chiral condensate at this point is found as
\beq
\sigma=0.822\,h^{1/5}(\equiv\sigma_0).
\eeq
The dotted line starting from the CP expresses the pseudocritical line
 determined by the location of the minimum in the sigma meson mass. 
The shaded area again represents the spinodal region, in which there is
 another state competing with the ground state.

\begin{figure*}[t]
\begin{center}
\includegraphics[width=0.62\textwidth]{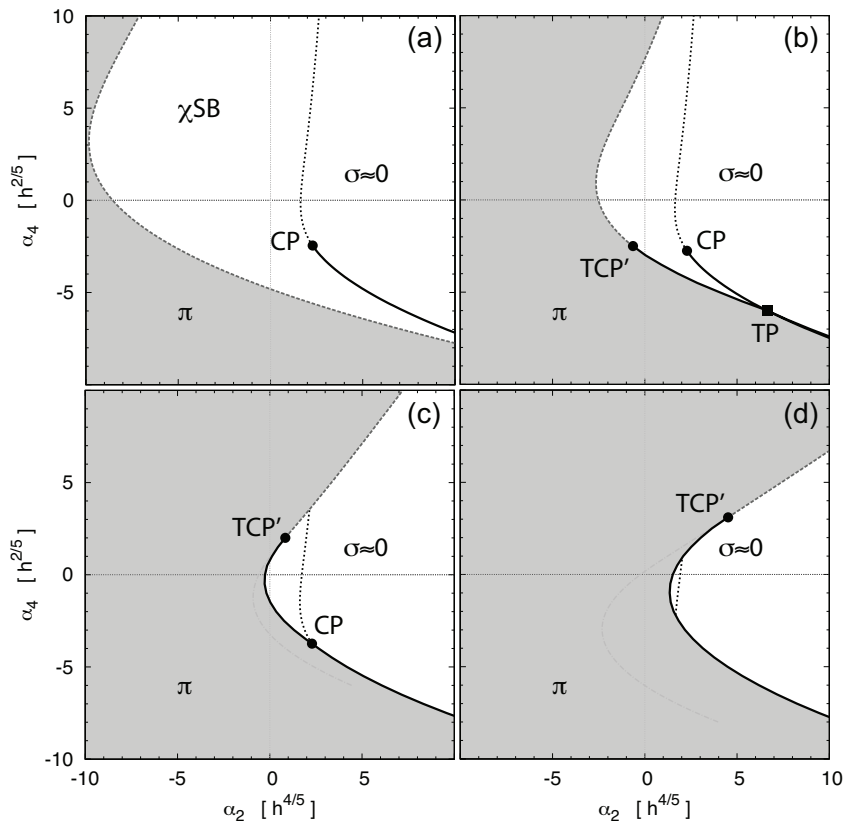}
\caption{The GL phase diagrams for several values of $\muI$;
from (a) to (d), $\muI$ increases as
 (a)~$\muI^2=0.2h^{2/5}$, 
 (b)~$\muI^2=0.5h^{2/5}$,
 (c)~$\muI^2\equiv\muIcs$, 
 and (d)~$\muI^2=3.0h^{2/5}$. For $\muIc$, see the text.
}
\label{fig:final}
\end{center}
\end{figure*}

\vspace*{0.0ex}
\emph{GL phase diagram for quark matter with an isospin imbalance:~}
Now, we discuss the influence of $\muI$ on the phase diagram.
The phase diagrams for several finite values of $\muI$ are shown in
 Fig.~\ref{fig:final}.
From Fig.~\ref{fig:final}(a) to Fig.~\ref{fig:final}(d), the value of
 $\muI$ increases.
We now suppress the spinodal region surrounding the line of the
 first-order chiral phase transition.

Let us start with (a), where the value of $\muI$ is set to
 $\muI^2=0.2h^{2/5}$, that is, in terms of $\sigma_0$,
 $\muI=0.54\sigma_0$.
In this case we notice that the structure in the proximity of the CP is
 unaffected even though the location of the CP is shifted downward
 according to
\beq
(\alpha_2^{\mathrm{CP}},\alpha_4^{\mathrm{CP}})%
 \to(\alpha_2^{\mathrm{CP}},\alpha_4^{\mathrm{CP}}-\muI^2).
\label{eq:shift}
\eeq
This can be easily understood from the coefficient of the $\sigma^4$
 term in Eq.~(\ref{eq:GLpot}).
Recalling the direction in which the local coordinate $\alpha_4$ points
 in the $(\mu,T)$ plane (Fig.~\ref{fig:map}), we expect that the CP
 shifts to the direction of lower temperature and higher chemical
 potential due to the isospin asymmetry.
This is consistent with the analysis done in Ref.~\cite{Ohnishi:2011jv}. 
Moreover, a recent analysis within a specific model shows at some
 critical value of $\muI$ the TCP (CP) can even touch the $\mu$ axis
 disappearing eventually from the phase diagram \cite{Ueda:2013sia}.
Apart from this simple shift of chiral phase transition, we notice that
 the continent of the PIC dominates the region deep in the $\chi\gr{SB}$. 
The PIC and $\chi\gr{SB}$ phases are separated by a second-order phase
 transition in all regions shown in the figure.

Figure \ref{fig:final}(b) shows how the situation changes when $\muI$ is
 increased to $\muI^2=0.5h^{2/5}$, that is, $\muI\cong0.86\sigma_0$.
We notice that the CP moves further downward according to
 Eq.~(\ref{eq:shift}), and the continent of the PIC gets wider
 as expected.
Moreover, the transition from the $\chi\gr{SB}$ to the PIC now has a
 branch of the first-order phase transition, which is drawn by a solid
 line. 
Accordingly, there appears a new tricritical point denoted by TCP$^\prime$
 on the critical line. 
This is actually the tricritical point at which three
 critical lines meet up, once we introduce an external field for the
 charged pion condensate.
The line for the first-order phase transition departing from
 the TCP$^{\prime}$ encounters the line of the first-order chiral phase
 transition at the point ``TP'', which stands for the triple point.
At the triple point, three phases---the $\chi\gr{SB}$, a nearly
 restored phase, and the PIC---coexist and compete.

Now, let us discuss Fig.~\ref{fig:final}(d) before Fig.~\ref{fig:final}(c). 
In Fig.~\ref{fig:final}(d), $\muI$ is set to a large value
 $\muI^2=3.0h^{2/5}$, which corresponds to $\muI\cong 2.1\sigma_0$. 
In this case, the PIC dominates a major part of the phase
 diagram, and the structure of the chiral phase transition is now 
 completely hidden. 
There remains a dotted line outside the PIC, which is just the chiral
 crossover.
The transition from the PIC to the $\chi\gr{SB}$ phase becomes
 widely of first order, and, accordingly, the location of the
 TCP$^\prime$ is shifted upward.

Since the CP is completely hidden in Fig.~\ref{fig:final}(d), there
 should be a critical value of $\muI$ at which the CP vanishes from the
 phase diagram.
In Fig.~\ref{fig:final}(c), we show this situation. 
The critical chemical potential is $\muI^2=1.477h^{2/5}$, which
 translates into $\muI\cong1.48\sigma_0\equiv\muIc$.
As shown in the figure, the CP comes across the line expressing the
 first-order phase transition between the PIC and $\chi\gr{SB}$
 phases.
This means when $\muI$ becomes large, the chiral critical point could
 fade out from the QCD phase diagram. 
The ratio of critical $\muI$ to $\sigma_0$, the chiral condensate at the
 CP, can be numerically evaluated as $\sim1.48$, which is universal
 being independent of any GL parameters to this order.

\section{Conclusion}\label{sec:conc}
We performed a systematic GL analysis on the effect of isospin asymmetry
 on the chiral crossover, the CP and its neighborhood.
We first focused on how the crossover is affected by the isospin
 density. 
To this aim, we derived a general GL potential up to the quartic order
 in two chiral four-vectors, $\phi$ and its parity partner $\varphi$.
Making use of the quark loop approximation together with a perturbative
 expansion in $\muI$, we have studied not only the nature of the phase
 transition to the PIC but also how it is affected by
 the effect of the $\gr{U}(1)_{\gr{A}}$ anomaly.
We found the effect of the isospin-flip odd $\lambda$ term in the
 potential makes the phase transition to first order at large $\lambda$. 
Since $\lambda$ vanishes at $\mu=0$ and increases with $\mu$, this may
 explain why the transition to the PIC is observed to be first order at
 finite $\mu$ in several model analyses
 \cite{Klein:2003fy,Barducci:2004tt}.
The effect of flavor mixing due to the $\gr{U(1)_A}$ anomaly was found
 to diminish the effect of the $\lambda$ term by locking two condensates,
 $\sigma_u$ and $\sigma_d$.
We have derived three model-independent universal
 ratios---$\muIc/\sigma_{\gr{pc}}$ at $\lambda=0$, 
 $\lambda^{\gr{TCP}}/\sigma_{\gr{pc}}$ and
 $\muI^{\gr{TCP}}/\sigma_{\gr{pc}}$ at the TCP---, which are independent of
 any GL parameters to the fourth order.

We then extended the analysis up to the sixth order of GL expansion,
 so as to study the isospin effect on the CP.
Restricting the analysis to the case with strong $\gr{U(1)_A}$ symmetry
 breaking, we studied how the CP and its neighborhood are affected by
 the incorporation of isospin density.
We found that it has remarkable effects; it not only causes
 a shift of the location of the CP, but also brings about the
 development of a sizable region for the homogeneous pion condensate. 
This leads to the appearance of new tricritical and triple points.
Moreover we showed that the CP disappears once the isospin chemical
 potential is increased above a critical value.
We derived the critical value $\muIc$ and a universal relation relating
 it with the size of chiral condensate at the CP, $\sigma_0$.

There are several directions into which the current work can be extended.
First, we need to take into account the possibility of
 inhomogeneous phases since they are known to play an important role
 near the TCP/CP
 \cite{Nakano:2004cd,Nickel:2009ke,Nickel:2009wj,Carignano:2010ac,%
Abuki:2011pf,Carignano:2012sx,Fukushima:2012mz}.
This is actually now under investigation \cite{Abuki}.
Second, the extension to three flavors would be interesting.
This would require the incorporation of a kaon condensate and a diquark
 condensate of the color-flavor locked type.
In particular, it is known that the interplay between the chiral
 and diquark condensates via the axial anomaly leads to a rich
 variety of phases and an appearance of new multicritical points
 \cite{Hatsuda:2006ps,Basler:2010xy}.
Third, the effect of vector interaction should be taken into account
 \cite{Kitazawa:2002bc,Zhang:2008wx}. 
This is especially needed when we look at dynamical aspects of the
 critical behavior.
In fact the dynamic universality class of the chiral CP is known to be
 the same as the liquid-gas CP \cite{Fujii:2003bz,Son:2004iv}.
Lastly, it is strongly desirable to seek the stiff experimental
 signatures of critical points observed here, such as those discussed
 for CP \cite{Stephanov:1998dy}.

\vspace*{0ex}
The author thanks K.~Suzuki for several useful comments.
A part of numerical calculations was carried out on SR16000 at YITP in
Kyoto University.

\end{document}